\documentclass{llncs}
\usepackage{amssymb}
\usepackage{amstext}
\usepackage{epsfig}
\usepackage{makeidx}  
\usepackage{url}

\begin{document}

\begin{frontmatter}

\pagestyle{headings}  

\mainmatter
\title{Pseudo-Random Bit Generation based on 2D chaotic maps of logistic type and 
its Applications in Chaotic Cryptography}

\titlerunning{PRBG based on 2D chaotic maps of logistic type}  

\author{C. Pellicer-Lostao \and R. L\'{o}pez-Ruiz}

\authorrunning{Pellicer-Lostao and L\'{o}pez-Ruiz  }   

\institute{Department of Computer Science and BIFI, \\
Universidad de Zaragoza, 50009 - Zaragoza, Spain,\\
\email{carmen.pellicer@unizar.es,}
\email{rilopez@unizar.es} }

\maketitle              

\begin{abstract}
Pseudo-Random Bit Generation (PRBG) is required in many aspects of cryptography as well as in other applications
of modern security engineering. In this work, PRBG based on 2D symmetrical chaotic mappings of logistic type is
considered. The sequences generated with a chaotic PRBG of this type, are statistically tested and the
computational effectiveness of the generators is estimated. Considering this PRBG valid for cryptography, the
size of the available key space is also calculated. Different cryptographic applications can be suitable to this
PRBG, being a stream cipher probably the most immediate of them.

\end{abstract}

\end{frontmatter}

\section{Introduction}
Many aspects of cryptography and modern security engineering depend upon the generation of pseudo-random
numbers. Examples are the use of nonces in authentication protocols, salts in certain signature
schemes, generation of keys or the keystreams of stream cyphers ~\cite{anderson},~\cite{menezes}.
The requirements of randomness in these generators vary according to their application. For example, the
generation of master keys normally requires high quality or entropy ~\cite{kocarev2}. But in the case of
nonces, uniqueness can be the main requirement for some protocols~\cite{anderson}.

Pseudo-Random Bit Generators (PRBG) are implemented by deterministic numeric algorithms and they should pass
several statistical tests ~\cite{menezes},~\cite{nist}, to prove themselves useful. These tests can be set up to
different levels of requirements of randomness depending on the future application of the PRBG. The security of
the entire cryptographic protocol or system, relies on the randomness quality of the generator
~\cite{kocarev2}.

Over the last two decades, several works have implemented PRBG based on chaotic systems (a complete survey can
be found in ~\cite{Li_thesis}). Chaotic systems have the property of beeing deterministic in the microscopic
space and behave randomly, when observed in a coarse-grained state-space. The sensitivity of chaotic maps to
initial conditions make them optimum candidates to relate minimal critical information about the input, in the
output sequence. Their iterative nature makes them fast computable and able to produce binary sequences with
extremely long cycle lengths.

In 2006 Madhekar Suneel proposes in ~\cite{suneel}, a method for pseudo-random binary sequence generation based
on the two-dimensional H\'{e}non map. The author also indicates that the choice of the H\'{e}non map is rather arbitrary
and that similar results should also be attainable with other 2D maps.

This paper explore precisely this possibility, and presents a finite automata scheme as the key to achieve that.
This comprehensive scheme is then applied to two particular 2D dynamical systems presented in ~\cite{ric}, which
are formed by two symmetrically coupled logistic maps. The pseudo-random properties of the generators obtained
that way, are investigated.

The chaotic PRBG algorithm described in this paper can be used in different ways. One of its applications, and
maybe the most immediate, could be the construction of practical stream ciphers. In this way, the chaotic PRBG
can expand a short key into a long keystream, which directly exclusive-or'ed with the clear text or message,
gives the ciphertext. The evaluation of the potential size of the key space and the computational cost of the
algorithm makes it worth to be considered.

The paper is structured as follows: Section 2 introduces chaotic PRBG. Section 3 describes statistical testing
of random sequences. Section 4 obtains a PRBG based on a 2D symmetrical chaotic map of logistic type. In Section
5 several binary sequences are obtained and tested. The computational cost of the PRBG algorithm and the size of
the key space for cryptographic applications are evaluated in Section 6. Section 7 remarks the final conclusions
and discusses further work to be done.

\section{Chaotic Random Bit Generation}
The inherent properties of chaos, such as ergodicity and sensitivity to initial conditions and control
parameters, connect it directly with cryptography characteristics of confusion and diffusion ~\cite{alvarez}.
Additionally chaotic dynamical systems have the advantage of providing simple computable deterministic
pseudo-randomness.

As a consequence of these observations, several works were presented since 1990s implementing PRBG based on
different chaotic systems ~\cite{suneel}, ~\cite{protopopescu}, ~\cite{kocarev2}, ~\cite{Lee}, ~\cite{Li_CCS}.
In some way, it could be said that chaos has brought into being a novel branch for PRNG in cryptography, called
chaotic PRNG.

An N-dimensional deterministic discrete-time dynamical system is an iterative map $f:\Re^N\rightarrow\Re^N$   of
the form:

\begin{equation}
X_{k+1} = f (X_k)
\end{equation}

where $k = 0,1\ldots n$. is the discrete time and $X_0,X_1\ldots X_n$, are the states of the system at different
instants of time.

In these systems, the evolution is perfectly determined by the mapping $f:\Re^N\rightarrow\Re^N$ and the initial
condition $X_{0}$. Starting from $X_{0}$, the \textit{initial state}, the repeated iteration of (1) gives rise
to a fully deterministic series of states known as an \textit{orbit}.

To build a chaotic PRBG is necessary to construct a numerical algorithm that transforms the states under
chaotic behaviour of the system into binary numbers. The existing designs of chaotic PRNGs use different
techniques to pass from the continuum to the binary world ~\cite{Li_thesis}. The most important are:

\begin{enumerate}
    \item Extracting one or more bits from each chaotic state along chaotic orbits ~\cite{protopopescu}.
    \item Dividing the phase space into m sub-spaces (defined through $N=log_2(m)$ threshold values),
    and output a binary number $i={0,1,…,m-1}$ if the chaotic orbit visits the $i_{th}$ subspace
     ~\cite{suneel}, ~\cite{kocarev2}.
    \item Combining the outputs of two or more chaotic systems to generate the pseudo-random numbers
     ~\cite{Lee},~\cite{Li_CCS}.
\end{enumerate}

Discrete or digital chaos implemented on computers with finite precision is normally called ``pseudo chaos". In
pseudo chaos dynamical degradation of the chaotic properties of the system may appear, as throughout iterations
pseudo orbits may depart from the real ones in many different and uncontrolled manners ~\cite{Li_degrad}. Even
so, the above exposed techniques are capable of generating sequences of bits or binary numbers, which appear
random-like from many aspects.

One may also consider that using high dimensional chaotic systems could offer additional advantages. While less
known, these systems whirl many variables at any calculation and the periodic patterns produced by the finite
precision of the computer are more difficult to appear than in the low dimensional cases.

In this paper the technique of dividing the phase space is followed and applied on two symmetrical
two-dimensional (2D) chaotic maps of logistic type.

\section{Statistical Tests Suites}
In general, randomness cannot be mathematically proved. Alternatively, different statistical batteries of tests
are used. Each of these tests evaluates a relevant random property expected in a true random generator. To test
a certain randomness property, several output sequences of the generator under test are taken. As one knows a
priori the statistical distribution of possible values that true random sequences would be likely to exhibit, a
conclusion can be obtained upon the probability of the tested sequences to be random.

Mathematically this is done as follows ~\cite{nist}. For each test, a \emph{statistic variable} $X$ is specified
along with its correspondent \emph{theoretical random distribution function} $f(x)$. For non-random sequences,
the statistic can be expected to take on larger values, typically far-out in the tails of $f(x)$. A
\emph{critical value} $x_{\alpha}$ is defined for the theoretical distribution so that $P(X>x_{\alpha})=
\alpha$, that is called the \emph{significance level} of the test. In the same way, theoretically other
distribution functions and a $\beta$ value could be defined to assess non-random properties. But in practice, it
is impossible to calculate all the distributions that describe non-randomness, for there are an infinite number
of ways that a data stream can be non-random.

When a test is applied, the test statistic value $X_s$ is computed on the sequence being tested. This test statistic
value $X_s$ is compared to the critical value $x_{\alpha}$. If the test statistic value exceeds the critical
value, the hypothesis for randomness is rejected. The rejection is done with a $(100*\alpha)\%$ probability of
having FALSE POSITIVE error (this is called a \emph{TYPE I error}, where the sequence was random and is
rejected). Otherwise is not rejected (i.e., the hypothesis is accepted) with a probability of $(100*\beta)\%$ of
error (this is called \emph{TYPE II error} or FALSE NEGATIVE, the sequence was non-random and is accepted). In
consequence passing the test merely provides a probabilistic evidence that the generator produces sequences
which have certain characteristics of random sequences.

For a given application, the value of $\alpha$ must be selected appropriately. This is because if $\alpha$ is
too high TYPE I errors may frequently occur (respectively if $\alpha$ is too low the same will happen for TYPE
II errors). For cryptographic applications typical values of $\alpha$ are selected between the interval
$\alpha\epsilon[0.001, 0.01]$, which is also referred as a confidence level in the interval $[99.9\%, 99\%]$ for
the test. Unlike $\alpha, \beta$ is not fixed, for it depends on the non-randomness defects of the generator.
Nevertheless $\alpha,\beta$ and the size of the tested sequence (n) are related . Then for a given statistic, a
critical value and a minimum n should be selected to minimize the probability of a TYPE II error $(\beta)$.

There exist different well-known sources of test suites available in the Internet. In the present work,
Marsaglia's Diehard test suite (in ~\cite{marsaglia}) and NIST Statistical Test Suite (in ~\cite{nist}) were
selected, for they are very accessible and widely used. Table 1 lists the tests comprised in these suites.
\begin{table} [h]
\label{table1}
\begin{center}
\begin{tabular}{|c|c|c|}
   \hline
  Number & Diehard test suite & NIST test suite \\
  \hline
  1 & Birthday spacings & Frequency (monobit) \\
  2 & Overlapping 5-permutation  & Frequency test within a block \\
  3 & Binary rank test &  Cumulative sums\\
  4 & Bitstream & Runs\\
  5 & OPSO & Longest run of ones in a block\\
  6 & OQSO & Binary matrix rank\\
  7 & DNS & Discrete fourier transform\\
  8 & Count-the-1's test & Non-overlapping template matching\\
  9 & A parking lot & Overlapping template matching\\
  10 & Minimum distance & Maurer's universal statistical\\
  11 & 3D-spheres & Approximate entropy\\
  12 & Squeeze &  Random excursions\\
  13 & Overlapping sums & Random excursions variant\\
  14 & Runs & Serial\\
  15 & Craps & Linear complexity\\
  \hline
\end{tabular}
\end{center}
\caption{List of tests comprised in the Diehard and NIST test suites.}
\end{table}
In both suites, the test statistic value $X_s$ obtained in each test is used to calculate a p-value that
summaries the strength of evidence against the randomness of the tested PRBG.

\section{Pseudo-Random Bit Generation based on two-dimensional chaotic maps of logistic type}

In ~\cite{ric}, L\'{o}pez-Ruiz and P\'{e}rez-Garc\'{\i}a analyze a family of three chaotic systems obtained by coupling
two logistic maps. The focus here will be made on models (a) and (b), which will be called systems A and B:

\begin{equation}
\label{systems}
\begin{array}{l}
SYSTEM\;\; A:\\ T_A:[0,1]\times[0,1]\longrightarrow[0,1]\times[0,1]\\ \; \\
x_{n+1} =\lambda(3y_n+1)x_n(1-x_n)
\\ y_{n+1}=\lambda(3x_n+1)y_n(1-y_n)
\end{array}
\hskip 0.5 cm
\begin{array}{l}
 SYSTEM\;\; B:\\ T_B:[0,1]\times[0,1]\longrightarrow[0,1]\times[0,1]\\ \; \\
 x_{n+1} =\lambda(3x_n+1)y_n(1-y_n)
 \\ y_{n+1} =\lambda(3y_n+1)x_n(1-x_n)
\end{array}
\end{equation}

Amazingly, these systems show the following symmetry $T_A(x,y)=T_B(y,x)$, which implies that
$T_A^2(x,y)=T_B^2(x,y)$. From a geometrical point of view, both present the same chaotic attractor in the
interval $\lambda\in[1.032, 1.0843 ]$. The dynamics in this regime is particularly interlaced around the
saddle point $P4$, that plays an important role for our proposes:

\begin{equation}
\label{p4} P4=[P4_x, P4_y],\hskip 8mm  where \hskip 4mmP4_x=P4_y=\frac{1}{3}
\left(1+\sqrt{4-\frac{3}{\lambda}}\right).
\end{equation}

\begin{figure}[]
\centerline{\includegraphics[width=12cm]{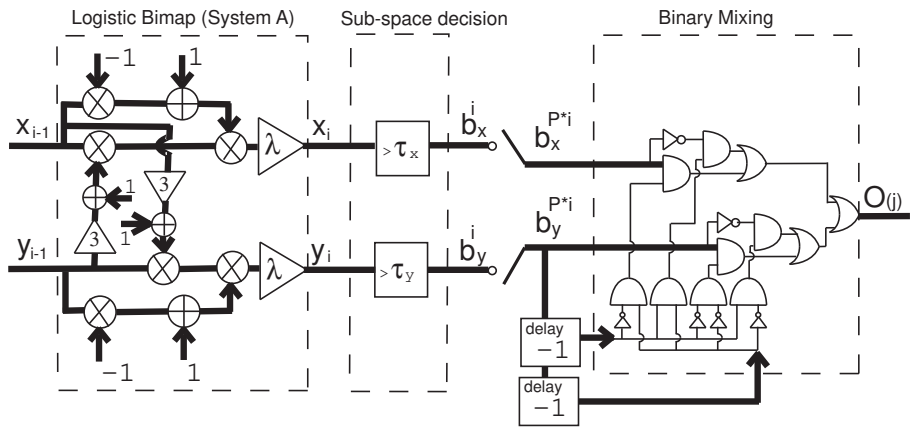}} \caption{Functional block structure of the proposed algorithm
in \cite{suneel} for System A.} \label{fig1}
\end{figure}

To obtain the Symmetric Coupled Logistic Map PRBG, the algorithm presented in ~\cite{suneel} is applied on
System A. Its functional block structure is represented in Fig.\ref{fig1} and it is explained in the following
paragraphs.

In this case, the technique of dividing the phase space in four sub-spaces is used. This is done in the block
named as \textit{Sub-space decision} in which the threshold values, $\tau_x$ and $\tau_y$, are
employed to convert the suite into a binary sequence, by means of the following equations:

\begin{equation}
b_x = \left\lbrace
\begin{array}{l}
0 \hskip 2mm {\rm if} \hskip 2mm x  \hskip 2mm \leq \hskip 2mm  \tau_x \\ 1  \hskip 2mm {\rm if} \hskip 2mm x
\hskip 2mm
> \hskip 2mm \tau_x
\end{array};
\right.
 \hskip 2cm
b_y =  \left\lbrace
\begin{array}{l}
0 \hskip 2mm {\rm if} \hskip 2mm y  \hskip 2mm \leq  \hskip 2mm \tau_y \\ 1  \hskip 2mm {\rm if} \hskip 2mm y
\hskip 2mm > \hskip 2mm \tau_y
\end{array}.
\right.
\end{equation}

A purely statistical procedure is proposed in ~\cite{suneel} to obtain $\tau_x$ (or $\tau_y$) as the median of a
large $T$ set of $x$ (or $y$) values. $\tau_x$ and $\tau_y$ are chosen after the first $T=1000$ iterations of
the system. After obtaining $S_x=\{b_x^i\}_{i=1}^{\infty}$ and $S_y=\{b_y^i\}_{i=1}^{\infty}$, they are sampled
with a frequency of $1/P$ (each $P$ iterations) and $B_x=\{b_x^{P*i} \}_{i=1}^{\infty}$ and
$B_y=\{b_y^{P*i}\}_{i=1}^{\infty}$ are obtained. The effect of skipping P consecutive values of the orbit is
necessary to get a random macroscopic behaviour. With this operation, the correlation existing between
consecutive values generated by the chaotic system is eliminated, in a way such that over a $P_{min}$, sequences
generated with $P>P_{min}$ will appear macroscopically random. Although $P$ is normally introduced as an
additional key parameter in pseudo-random sequences generation ~\cite{kocarev3}, it strongly determines the
speed of the generation algorithm, so it is recommended to be kept as small as possible.

The output binary pseudorandom sequence $O(j)$ is obtained by a mixing operation of the actual and previous
values of the sequence $B(j)=[B_x(j),B_y(j)]$ given by the truth table sketched in Table 2.

\begin{table}
\label{table2}
\begin{center}
\begin{tabular}{|c|c|c|}
  \cline{2-3}
  \multicolumn{1}{c|}{} & \multicolumn{2}{|c|}{$B_y(j-1)$}\\
   \hline
  $B_y(j-2)$ & 0 & 1\\
  \hline
   0 & $B_x(j)$ & ${\rm Not}(B_x(j))$\\
   \hline
   1 & $B_y(j)$ & ${\rm Not}(B_y(j))$\\
  \hline
\end{tabular}
\end{center}
\caption{Truth table generating the binary sequence.}
\end{table}

\begin{figure}[h]
\centerline{\includegraphics[width=3.5cm]{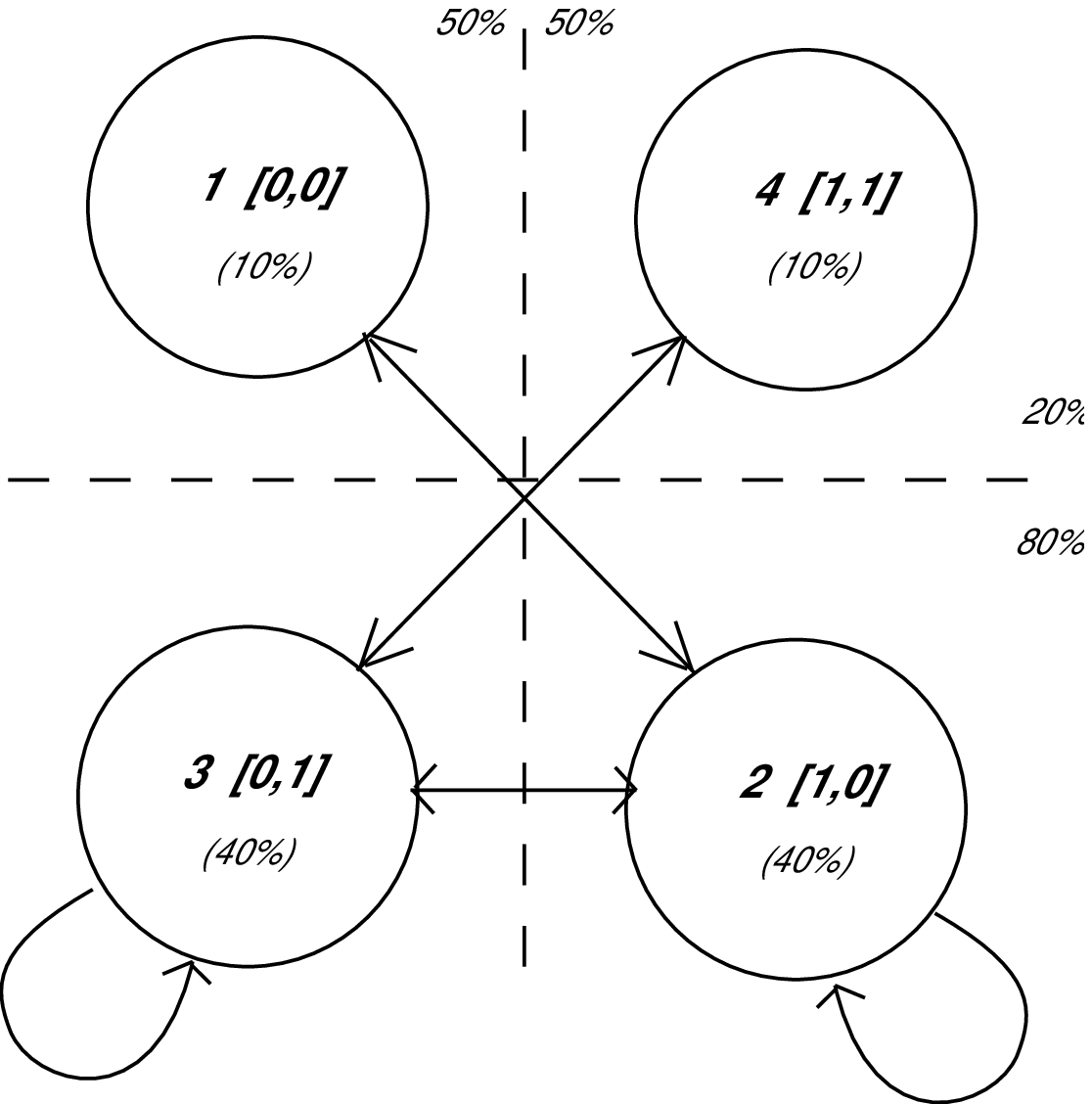}\hskip 5mm\includegraphics[width=5cm]{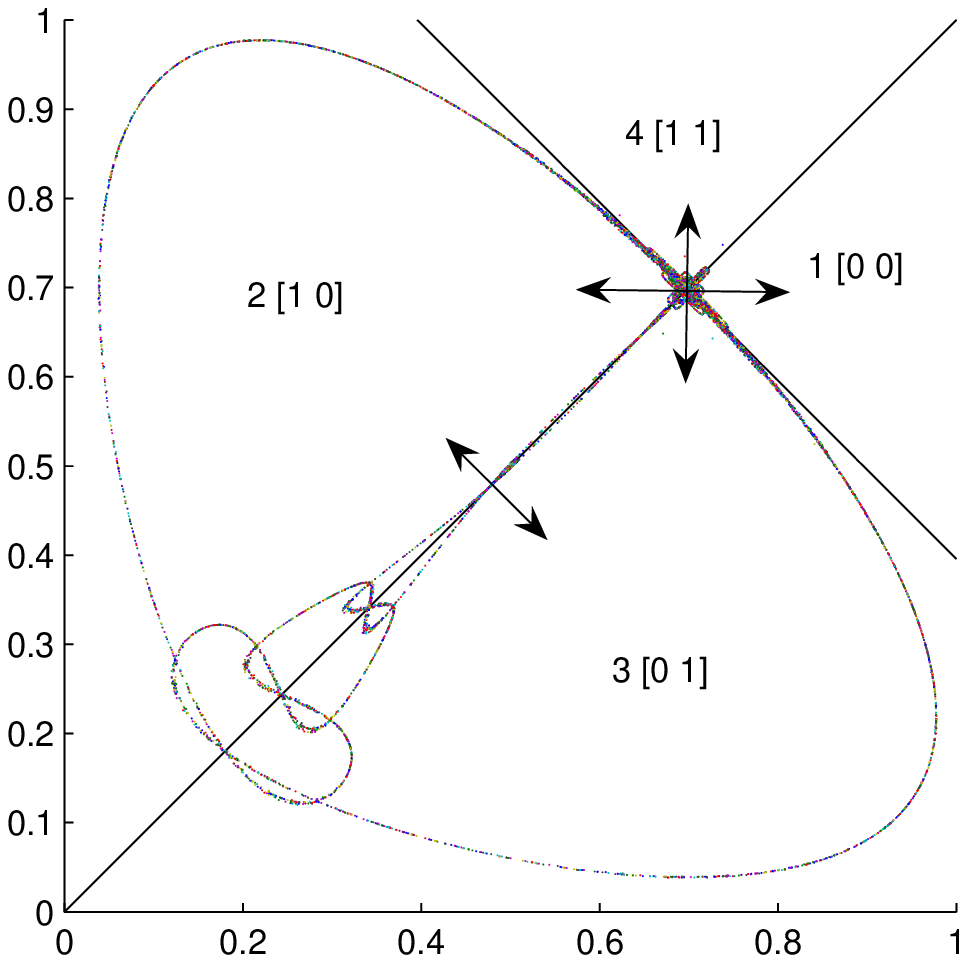}}
\centerline{(a)\hskip 5cm (b)} \centerline{\includegraphics[width=3.5cm]{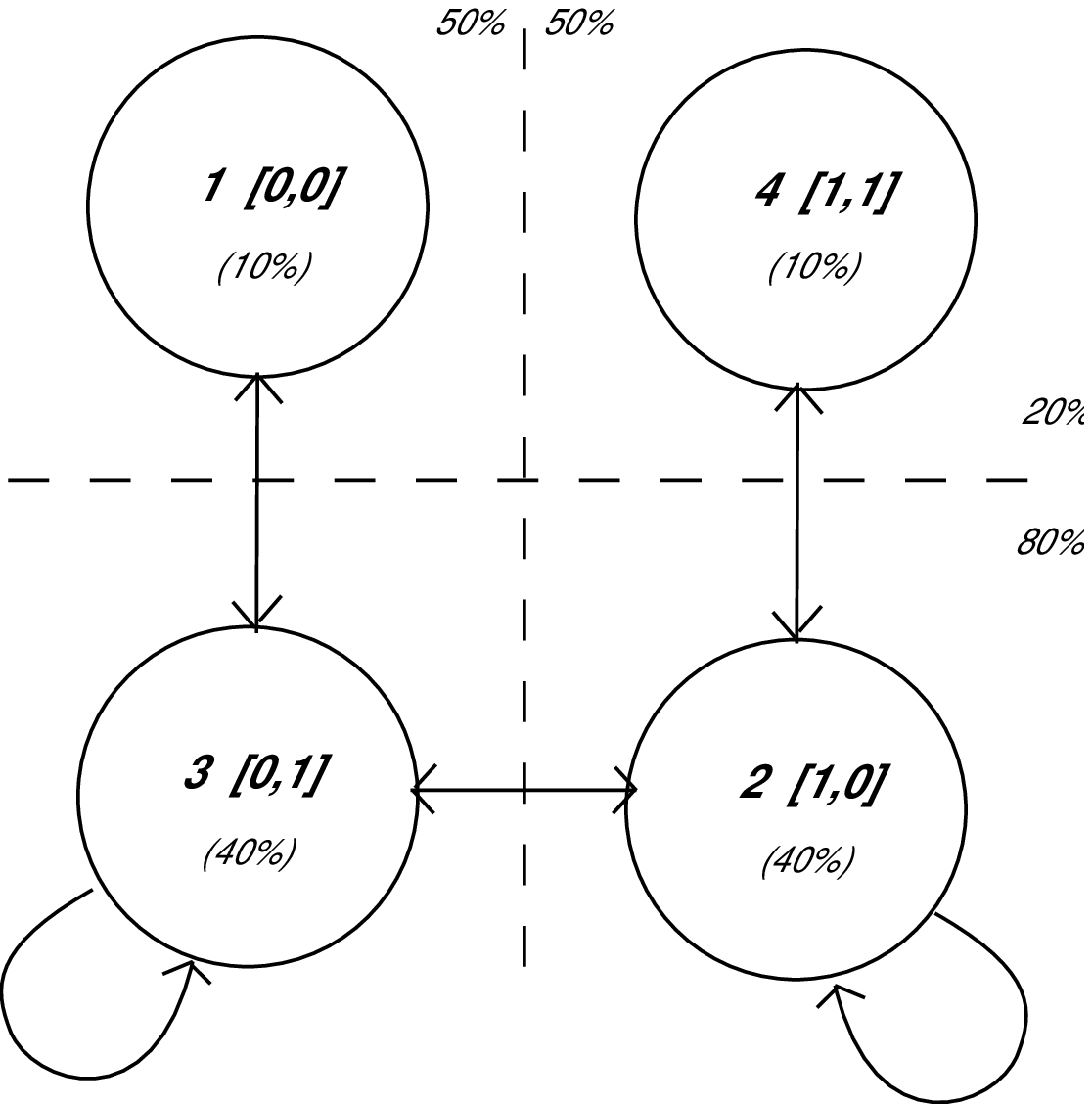}\hskip
5mm\includegraphics[width=5cm]{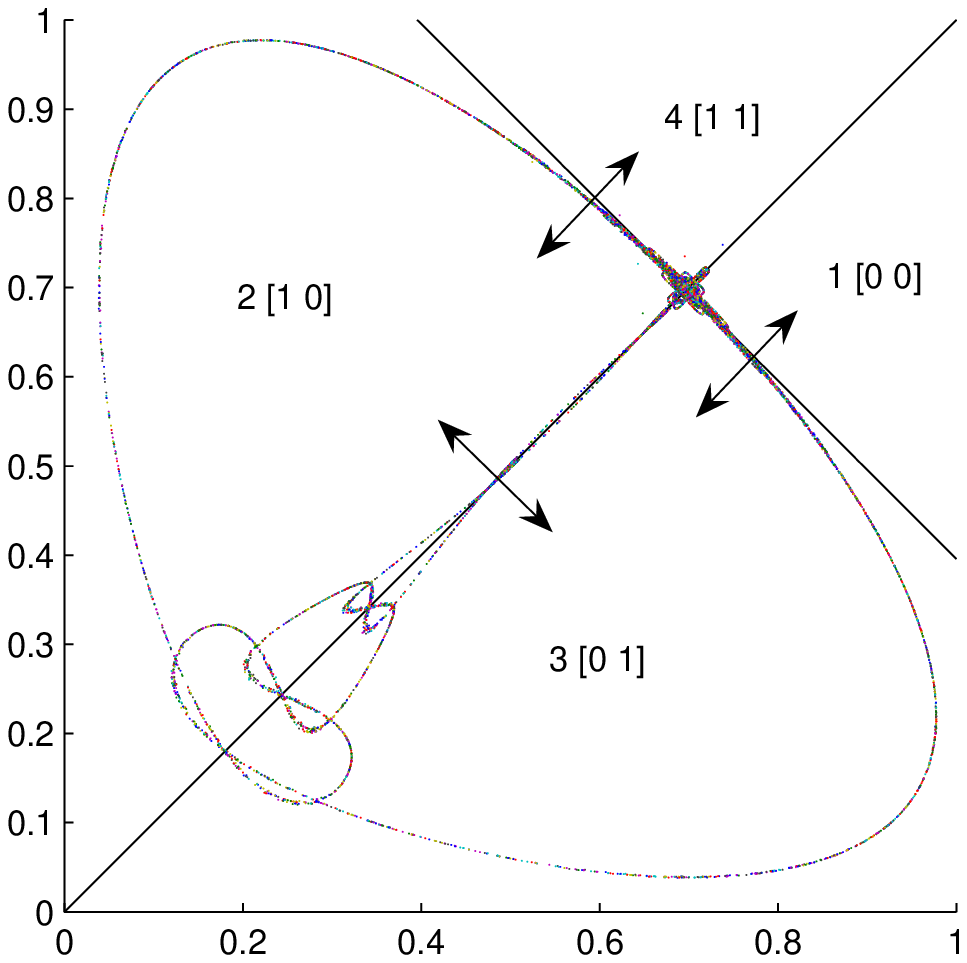}} \centerline{(c)\hskip 5cm (d)} \caption{(a) Finite automata and (b)
final sub-space division for System A. (c) Finite automata and (d) final sub-space division
for System B. (In both cases, $\lambda= 1.07$).} \label{fig2}
\end{figure}

Unfortunately the sequences so formed do not pass the minimum requirements of randomness assessed by Diehard
test suite. At this point, it is noticed that, to obtain good results, the geometrical characteristics of the
system must be taken into account. More precisely, it is found out that the division of the phase space in four
sub-spaces must be defined, in a way that the system visits each sub-space according to a particular finite
state automata. This automata is inferred from the H\'{e}non map behaviour in ~\cite{suneel} and it has a particular
pattern of visits of the four sub-spaces. For Systems A and B, this finite automata is depicted in Fig.
\ref{fig2}(a) and \ref{fig2}(c).

Let us name the sub-spaces corresponding to $[b_x^i, b_y^i]$ with values $[0,0]$, $[1,0]$, $[0,1]$ and $[1,1]$ as
1,2,3 and 4. Although the four sub-spaces are not visited equally, there exists a symmetry of movements between
sub-spaces 1-3 and 2-4, which has a characteristic mixing of 50\% and 50\%, as long as a predominant (80\%) and
constant transition between 3 and 2. This leads to a highly variation of binary values in sequences $S_x$,
$S_y$. In the end, these conditions give the final result of an output sequence $O(j)$ with a proper balance of
zeros and ones, or put it in another way, with pseudo-random properties.

To get this automata for the symmetric coupled logistic maps Systems A and B, one should chose the diagonal
axis, which divides phase space in two parts, each of which is equally visited (50\%). And additional
statistical calculus is required to divide these two sub-spaces, in another two with a visiting rate of
40\%-10\% each one. When this is done, one can observe that this is got by merely selecting P4 and the line
perpendicular to the axis in P4 as the other division line. The final sub-space division for each system  is
presented in Fig. \ref{fig2}(b) and \ref{fig2}(d), along with the indications of the evolution of the visits to
each sub-space.

\begin{figure}[h]
\centerline{\includegraphics[width=12cm]{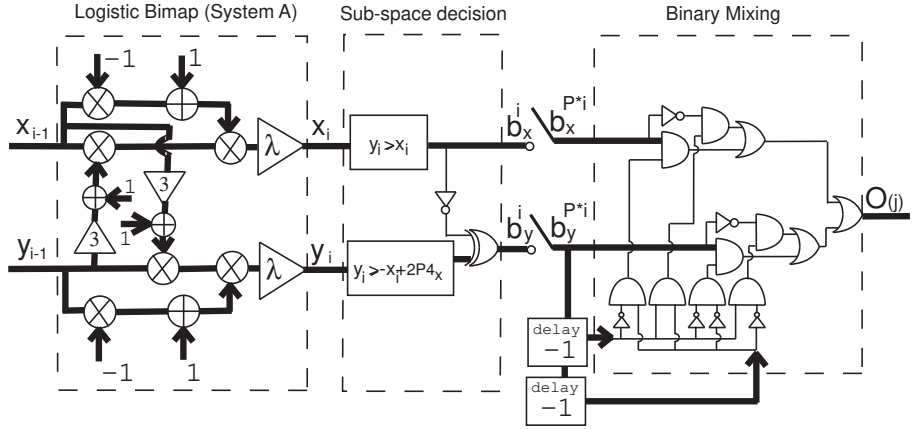}} \caption{Functional block structure of the PRBG applied to the
symmetric coupled logistic map PRBG with System A.} \label{fig5}
\end{figure}

Finally the initial algorithm in Fig. \ref{fig1} applied to System A, is modified with the appropriate sub-space
decision block. The final PRBG functional scheme is represented in Fig. \ref{fig5}.

As a direct application to cryptography, the PRBG could be used for the construction of a stream cipher.
Different initial conditions, $x_0$ and $y_0$ and parameters $\lambda$ and $P$, can be applied to the input of
the system and be used as a key to generate the keystream, or output sequence $O(j)$. The keystream $O(j)$ can
be XORed directly with a clear text obtaining that way the ciphertext.

 Different sequences are obtained with the system of Fig. \ref{fig5} in next section. Their randomness is assessed
 and demonstrates them statistically valid for cryptographic applications. This may indicate that the automata
scheme presented in Fig. \ref{fig2}(a) and \ref{fig2}(c) represents a sufficient condition to obtain
pseudo-randomness. Consequently, it can represent a systematic scheme to extend the algorithm in ~\cite{suneel}
to get PRBG on other chaotic maps.

\section{Pseudo-Random Sequences Statistical testing}
To assess the randomness of the PRBG obtained in the previous section with systems A and B, several sequences
are obtained and submitted to the Diehard ~\cite{marsaglia} and NIST ~\cite{nist} test suites described in
section 3. Similar results were found for both systems and for simplicity, only those obtained with system A
will be presented here after. Ten sequences were generated with six different sets of initial conditions. Their
characteristics are described in Table 3.

\begin{table} [h]
\label{table3}
\begin{center}
\begin{tabular}{|c|c|c|c|c|c|c|}
  \hline
  Sequence & S1 & S2 & S3 & S4 & S5 & S6\\
  \hline
  $x_0$ & $0.989125$ & $0.491335$ & $0.672757$ & $0.726874$ & $0.39565$ & $0.999851$\\
  $y_0$ & $0.689125$ & $0.691335$ & $0.497757$ & $0.901874$ & $0.49565$ & $0.649851$\\
  $\lambda$ & $1.04869$ & $1.05392$ & $1.06961$ & $1.08007$ & $1.06438$ & $1.07489$\\
  $P_{Dmin}$& $55$ & $45$ & $35$ & $47$ & $n.a.$ & $n.a.$ \\
  $P_{Nmin}$& $83$ & $105$ & $83$ & $83$ & $100$ & $85$ \\
  \hline
\end{tabular}
\end{center}
\caption{Parameters $P_{Dmin}$ and $P_{Nmin}$ for different sequences $S_i$, $i=1,..,6$, with different initial
conditions $(x_0,y_0)$ and map parameter $\lambda$.}
\end{table}

Six of them (S1,S2,S3,S4,S5 and S6) were tested with Nist tests suite with 200 Mill. of bits and four of them
(S1,S2,S3 and S4) were tested with Diehard tests suite with 80 Mill. of bits. Here, the parameters $P_{Dmin}$
and $P_{Nmin}$ are the minimum sampling rate or shift factor, $P_{min}$, over which, all sequences generated
with the same initial conditions and $P>P_{min}$ pass Diehard or Nist tests suites, respectively. It is observed
here, that the Nist tests suite requires a higher value of $P_{min}$ and that S5 and S6 were not tested with
Diehard battery of tests.

In the Diehard tests suite, each of them returns one or several p-values which should be uniform in the interval
[0,1) when the input sequence contains truly independent random bits. The significance level of the tests was
set properly for cryptographic applications ($\alpha=0.01$). The software available in ~\cite{marsaglia} provide
a total of 218 p-values for 15 tests, and the uniformity requirement can be assessed graphically plotting them
in the interval [0,1).

\begin{figure}[h]
\centerline{\includegraphics[width=6cm]{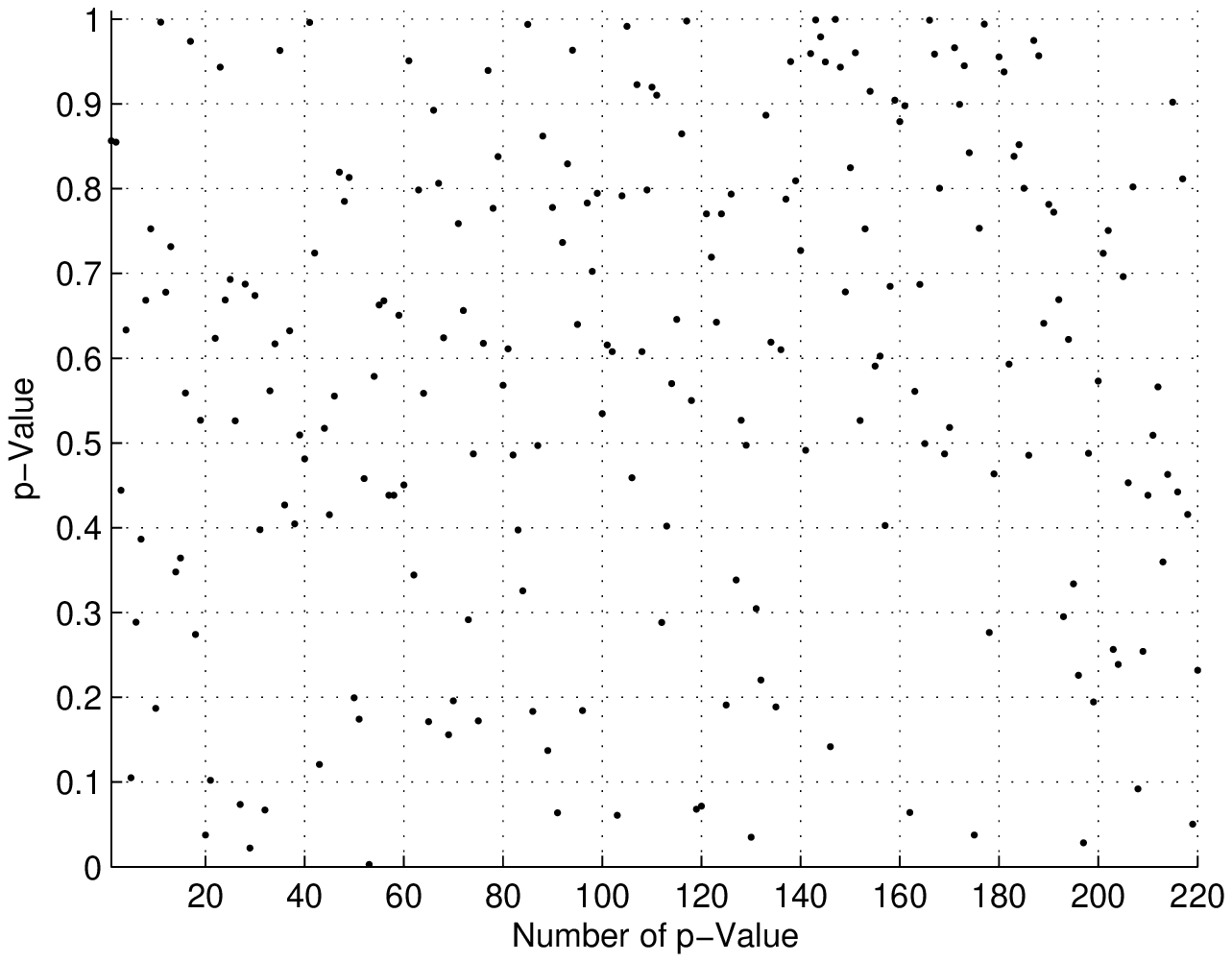}\hskip 5mm\includegraphics[width=6cm]{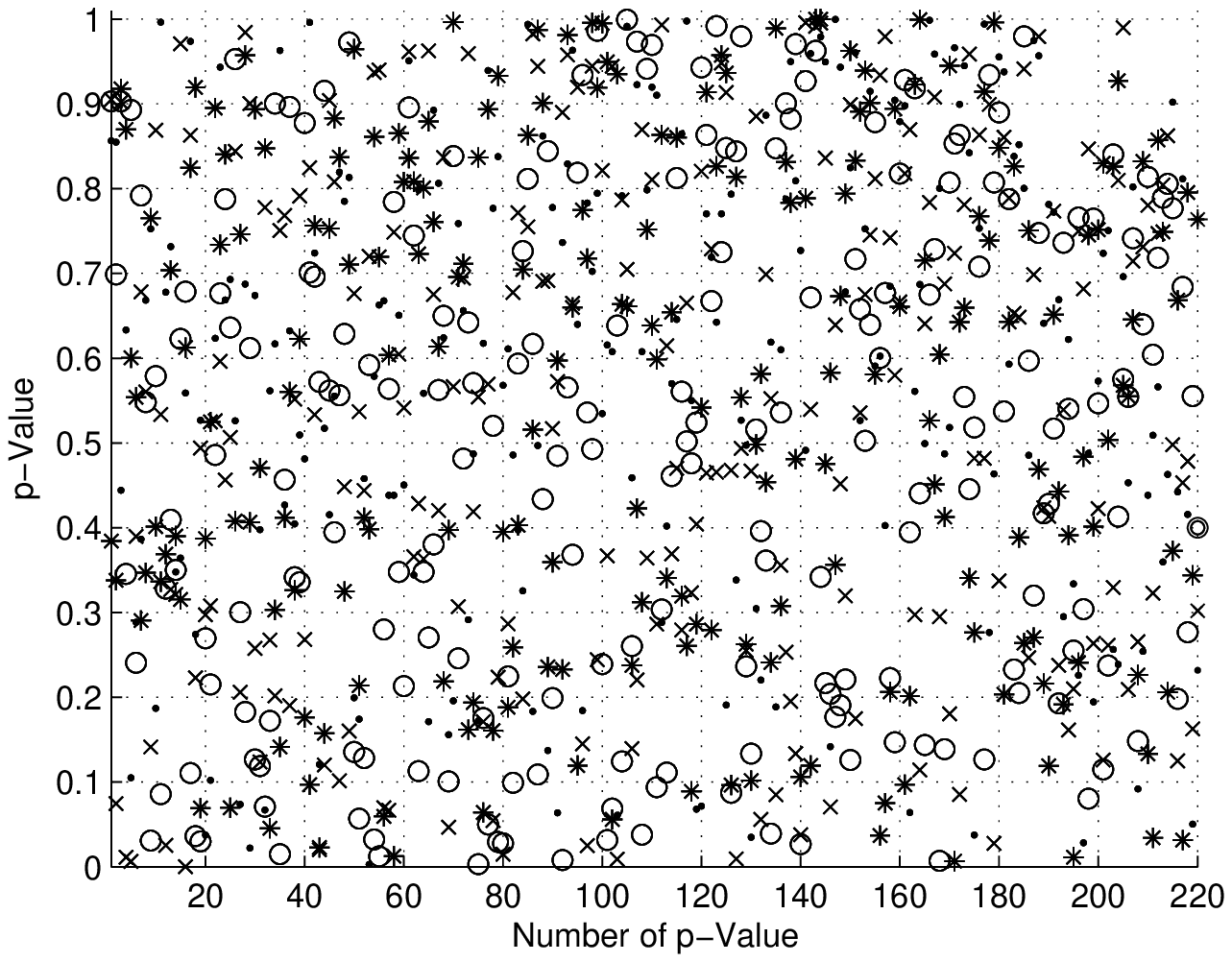}}
\centerline{(a)\hskip 6cm (b)} \caption{(a) Diehard test suite p-values obtained with all tests for initial
conditions S1 with $P=P_{Dmin}=55$. (b) p-values obtained for initial conditions
S1($\bullet$),S2($\circ$),S3($\ast$) and S4($\times$) with $P=P_{Dmin}$ value in Table 3.} \label{fig6}
\end{figure}

Fig. \ref{fig6}(a) shows the uniformity distribution of the p-values over the interval [0,1) obtained for a
sequence with initial conditions S1 and $P_{Dmin}=55$. Sequences with initial values S1 to S4 where proved to
pass the Diehard battery of tests. Fig. \ref{fig6}(b) presents a graphical representation of the p-values
obtained for each sequence with sampling factor $P=P_{Dmin}$ value in Table 3. It can be observed that some
p-values are occasionally near 0 or 1. Although it can not be well appreciated in the figure, it has to be said
that those never really reach these values.

In the Nist tests suite ~\cite{nist}, one or more p-values are also returned for each sequence under test. The
significance level of the tests was set to $\alpha=0.01$, as in the Diehard case. These tests require a
sufficiently high length of sequences and to prove randomness in one test, two conditions should be verified.
First, a minimum percentage of sequences should pass the test and second, the p-values of all sequences should
also be uniformly distributed in the interval $(0,1)$.

For this case, each of the six sequences with initial conditions S1 to S6 are arranged in 200 sub-sequences of
1Mill. bits each and submitted to the Nist battery of tests. Sequences $S$ proved to pass all tests over a
minimum value of $P_{Nmin}$, shown in Table 3.

\begin{figure}[]
\centerline{\includegraphics[width=6cm]{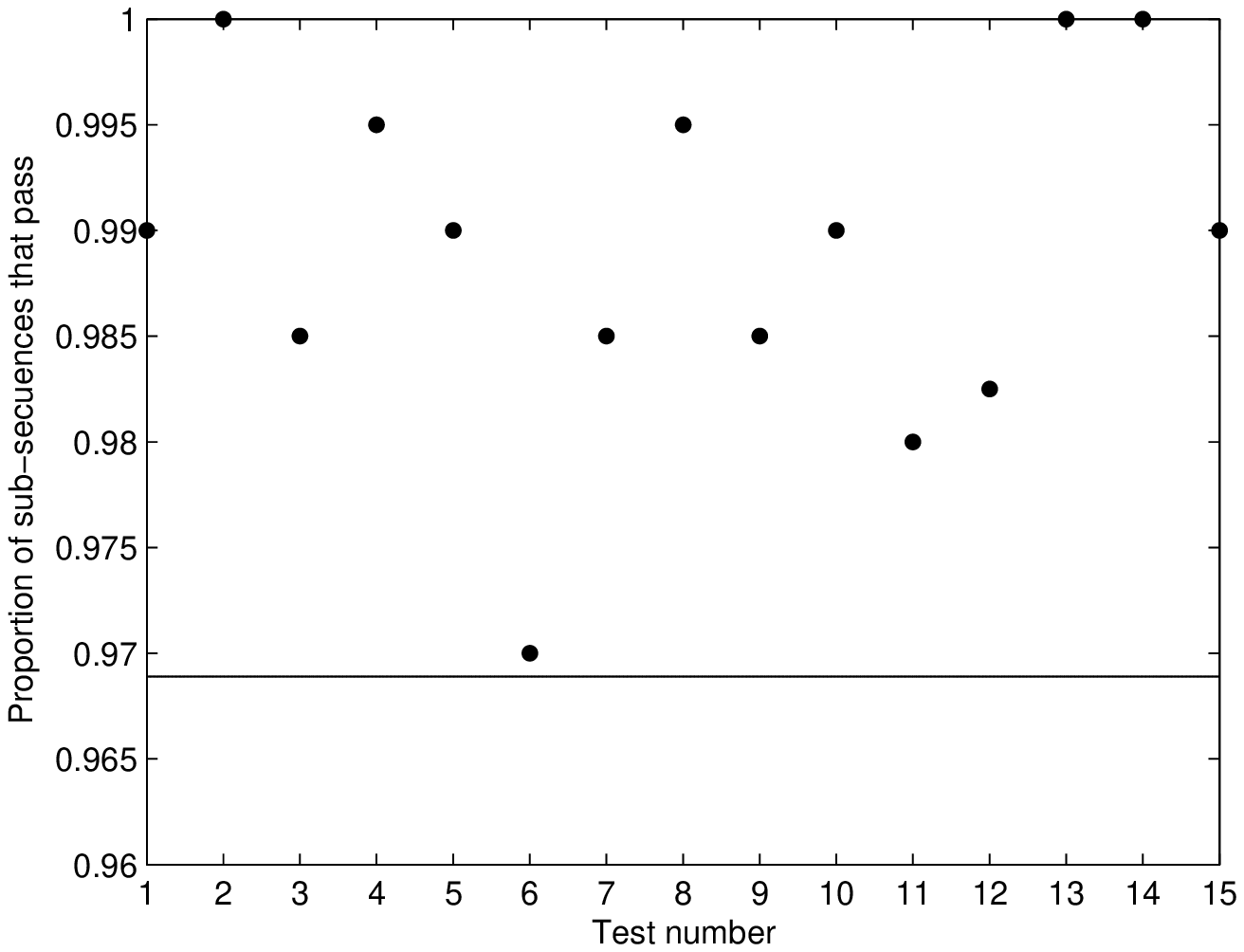}\hskip 0mm\includegraphics[width=6.5cm]{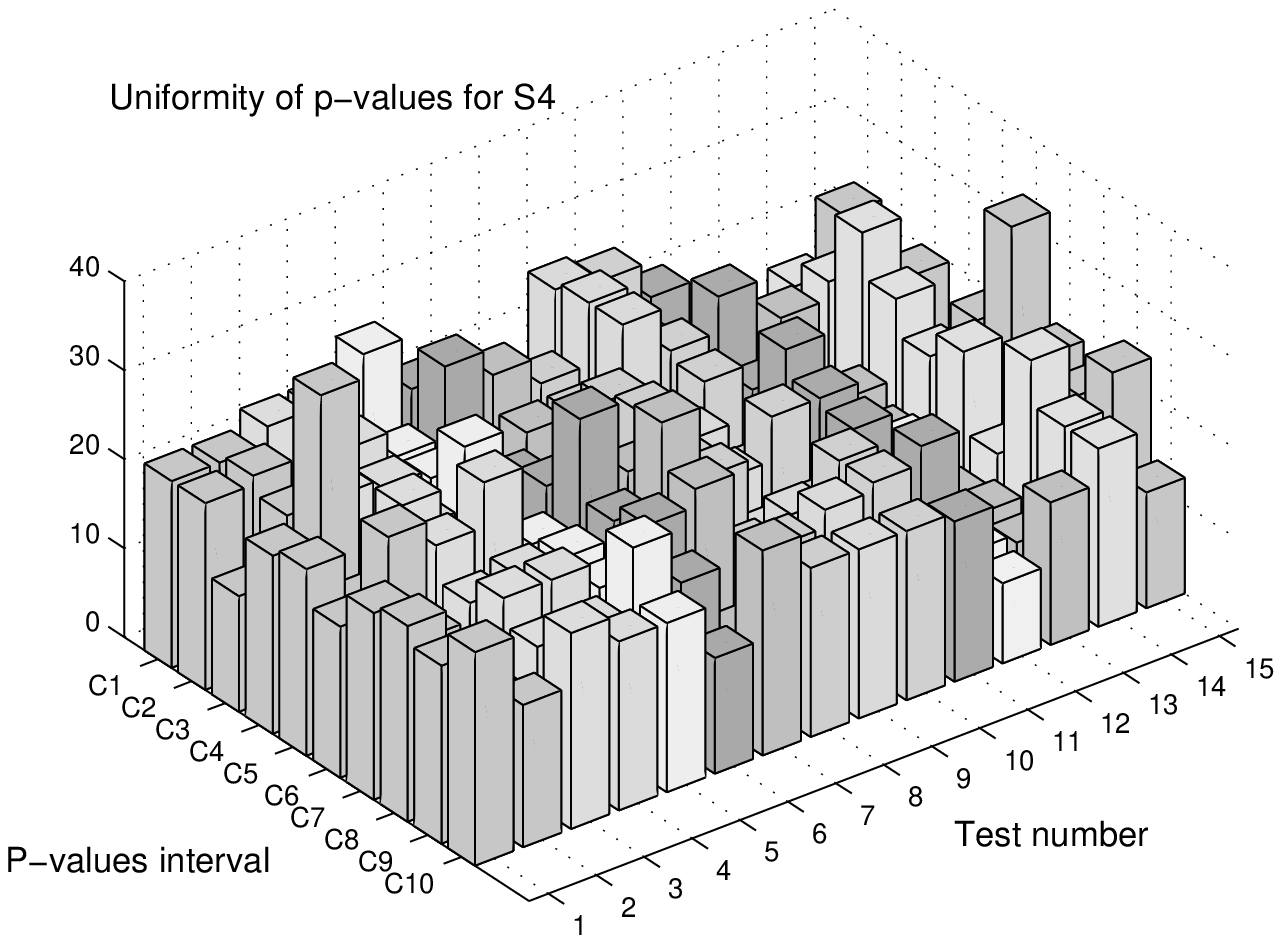}}
\centerline{(a)\hskip 5.5cm (b)}\caption{In (a), the proportion of sub-sequences S1 that passes each test is
displayed. In (b) The distribution of p-values of S4 is examined for each test to ensure uniformity. The
interval between 0 and 1 is divided in ten sub-intervals $(C1,C2,...,C10)$, and the p-values that lie within
each subinterval are counted and plotted.} \label{fig7}
\end{figure}

In Fig. \ref{fig7}(a) and \ref{fig7}(b), the results obtained for $S1$ and $S4$ respectively are graphically
presented, as an example of what was obtained for each $S$. The tests in the suite are numbered according to
Table 1. Fig. \ref{fig7}(a) represents the percentage of the 200 sub-sequences of S1, that pass each of the 15
tests of the suite. These percentages are over the minimum pass rate required of $96.8893\%$ for a sample size =
200 binary sub-sequences. Fig. \ref{fig7}(b) describes the uniformity of the distribution of p-values obtained
for the 15 tests of the suite. Here, uniformity is assessed. The interval (0,1) is divided in ten subintervals
$(C1,C2,...,C10)$ and the number of p-values that lay in each sub-interval, among a total of 200, are counted
and proved to be uniform.

\section{Key space size and computational cost}
To establish the complexity, and consequently the speed of the PRBG described in Fig. \ref{fig5}, the principle
of invariance is observed. This says that the efficiency of one algorithm in different execution environments
differs only in a multiplicative constant, when the values of the parameters of cost are sufficiently high.

In this sense, the asymptotic behaviour of the computational cost of the PRBG is governed by the calculus
performed in the chaotic block. This block performs P iterations to obtain an output bit, $O(j)$.

The capital theta notation ($\Theta$) can be used to describe an asymptotic tight bound for the magnitude of
cost of the PRBG. And consequently, the 2D symmetric coupled logistic maps have a computational cost or
complexity of order $\Theta(P*n)$.

When considered for cryptographic applications, the key space is determined by the interval of the parameter
$\lambda$ and the initial conditions that keep the dynamical system in the chaotic regime. These are
$\lambda\in[1.032, 1.0843 ]$ , $x_0\in(0, 1)$ and $y_0\in(0, 1)$. The sampling parameter can also be considered
as another parameter of the key space. One must observe that $P$ should be kept in a suitable range, so that the
PRBG is fast enough for its desired application. These intervals can be denoted with brackets and calculated as
$[\lambda]=0.0523$, $[x_0]=1$, $[y_0]=1$ and $[P]=4890$, when taking $[P]\in[110, 6000]$ as the range of the
sampling factor.

Let us consider $\epsilon_{32}\thickapprox1.1921\times10^{-7}$ as the smallest available precision for
fixed-point representation with 32 bits and its correspondent magnitude
$\epsilon_{64}\thickapprox2.2204\times10^{-16}$ for floating-point numbers with 64 bits. These quantities give
us the maximum number of possible values of every parameter in any of the two representations. This is easily
computed dividing the intervals by $\epsilon$, as $K_{\lambda}=[\lambda]/\epsilon$, $K_{x_0}=[x_0]/\epsilon$,
$K_{y_0}=[y_0]/\epsilon$ and $K_{P}=[P]/\epsilon$. The total size of representable parameter values is given by
$K$, calculated as $K= K_{\lambda}\times K_{x_0}\times K_{y_0}\times K_{P}$. $K$ is the size of the available
key-space and its logarithm in base 2 gives us the available length of binary keys to produce pseudo-random
sequences in the generator.

The values obtained for each number precision, are $K_{32}= 1.53\times10^{30}$ with a key length of 100 bits for
single precision and $K_{64}= 1.27\times10^{65}$, with a key length of 216 bits for double precision. These
results are encouraging for recommending the use of the PRBG in Fig. \ref{fig5} for cryptographic applications,
where a length of keys greater than 100 is considered strong enough against brute force attacks,
~\cite{alvarez}.

\section{Conclusions}
In the present work, a refinement of the algorithm exposed in ~\cite{suneel} by M. Suneel is presented. It
consists of the introduction of a finite automata that makes possible its application to other chaotic maps. In
some way, this finite automata could be said to extend the range of application of this algorithm for other 2D
chaotic systems. This is referred in ~\cite{Li_thesis} as making the PRBG chaotic-system-free.

The fact is that, while systematic, the scheme presented in this paper is not straight-forward. This is because
building the finite automata requires necessarily a detailed study of the geometrical properties of the
dynamical evolution of the chaotic system.

The authors apply this technique to build two new PRBG using two particular 2D dynamical systems formed by two
symmetrically coupled logistic maps. A set of different pseudo-random sequences are generated with one of the
PRBG. Statistical testing of these sequences shows fine results of random properties for the PRBG.

The estimation of the PRBG computational cost has an asymptotic tight bound of $\Theta(P*n)$. The available size
of the key space is also calculated and a minimum length of binary keys of 100 and 216 bits is obtained for
simple and double precision respectively. These preliminary results indicate a promising quality of the PRBG for
cryptographic applications.

Consequently, the chaotic PRBG algorithm could be of use for different applications in security engineering. A
direct application in cryptography could be the construction of a stream cipher. This can be easily obtained
when the output sequences $O(j)$ in Fig. \ref{fig5} are used as a keystream. Then, this can be directly XORed
with a clear text obtaining that way the ciphertext.

For future work, the authors plan to consider the geometrical properties of the logistic bi-mappings to enhance
the performance of the presented PRBG.


\begin{thebibliography}{99}


    \bibitem{anderson}
    Anderson, R.: Security Engineering, A Guide to Build Dependable Distributed Systems,
    http://www.cl.cam.ac.uk/~rja14/book.html
    John Wiley and Sons Inc., New York (2001)

    \bibitem{menezes}
    Menezes, A., van Oorschot, P., Vanstone, S.:  Handbook of Applied Cryptography.
    CRC Press, Florida (1997)

    \bibitem{nist}
    NIST Special Publication 800-22: A Statistical Test Suite for the Validation of Random Number Generators and
    Pseudo Random Number Generators for Cryptographic Applications, (2001)

    \bibitem{Li_thesis}
    Li, S.: Analyses and New Designs of Digital Chaotic Ciphers.
    PhD thesis, School of Electronic and Information Engineering, Xi'an Jiaotong University (2003)

    \bibitem{suneel}
    Madhekar, S.: Cryptographic Pseudo-Random Sequences from the Chaotic H\'{e}non Map.
    http://arxiv.org/abs/cs/0604018, (2006)

    \bibitem{ric}
    Lopez-Ruiz, R., P\'{e}rez-Garcia, C.: Dynamics Maps with a Global Multiplicative Coupling.
    Chaos, Solitons and Fractals, Vol.1, pp 511-528, (1991) ; Lopez-Ruiz, R., 
	Fournier-Prunaret, D.: Complex Patterns on the Plane: Different Types of Basin Fractalization 
	in a Two-Dimensional Mapping, International Journal of Bifurcation and Chaos, 
	Vol. 13, 287-310 (2003)

    \bibitem{alvarez}
    Alvarez G., Li, S.: Some Basic Cryptographic Requirements for Chaos-Based Cryptosystems.
    International Journal of Bifurcation and Chaos, Vol.16, pp 2129-2151 (2006)

    \bibitem{Li_degrad}
    Li, S., Chen,G., Mou, X.: On the dynamical degradation of digital piecewise linear chaotic maps.
    International Journal of Bifurcation and Chaos, Vol.15, pp 3119-3151 (2005)

    \bibitem{protopopescu}
    Protopopescu, V.A., Santoro, R.T., Tollover,J.S.: Fast secure encryption-decryption method 
	based on chaotic dynamics. US Patent No. 5479513 (1995)

    \bibitem{kocarev2}
    Stojanovski, T., Kocarev, L.: Chaos based Random Number Generatiors. Part I: Analysis.
    IEEE Transactions on Circuits and Systems I: Fundamental Theory and Applications, Vol.43, pp 281-288 (2001)

    \bibitem{Lee}
    Po-Han, L., Soo-Chang, P., Yih-Yuh, C.: Generating chaotic stream ciphers using chaotic systems.
    Chinese Journal of Physics, Vol.41, pp 559-581 (2003)

    \bibitem{Li_CCS}
    Li, S., Mou, X., Cai, Y.: Pseudo-random bit generator based on couple chaotic systems and its application in stream-ciphers cryptography.
    INDOCRYPT 2001, LNCS, Vol. 2247, pp 316-329, Springer, Heidelberg (2001).

    \bibitem{marsaglia}
    Marsaglia, G.: The diehard test suite, 1995.
    http://stat.fsu. edu/~geo/diehard.html

    \bibitem{kocarev3}
    Kocarev, L.: Chaos-based Cryptography: a brief overview.
    IEEE Circuits and Systems Magazine, Vol.1, pp 6-21 (2001)

\end{thebibliography}
\end{document}